\input amstex
\documentstyle{amsppt}
\magnification=1200
\pageheight{47.5pc}
\vcorrection{-0.5pc}
\NoBlackBoxes
\NoRunningHeads
\define\slt{\operatorname{sl}(2,\Bbb C)}

\define\G{\operatorname{G}}
\define\SUt{\operatorname{SU}(3)}

\define\PSL{\operatorname{PSL}}
\define\sLtwo{\operatorname{sl}(2,\Bbb C)}

\document\qquad\qquad\qquad\qquad\qquad\qquad\qquad\qquad\qquad
$\boxed{\boxed{\aligned
&\text{\eightpoint"Thalassa Aitheria" Reports}\\
&\text{\eightpoint RCMPI-95/05}\endaligned}}$\newline
\ \newline
\ \newline
\ \newline
\topmatter
\title Octonionic binocular mobilevision.\linebreak
An overwiew.
\endtitle
\author Denis V. Juriev\endauthor
\address\newline\rm
"Thalassa Aitheria" Research Center\newline
for Mathematical Physics and Informatics,\newline
ul.Miklukho-Maklaya 20-180, Moscow 117437 Russia
\endaddress
\email
denis\@juriev.msk.ru
\endemail
\abstract
This paper is a compact overview of the heuristic approach to the recently
elaborated octonionic binocular mobilevision [1].
\endabstract
\endtopmatter

\head 1. Interpretational geometries, anomalous virtual realities, and
mobilevision\endhead

\subhead 1.1. Interpretational geometry \endsubhead

A geometry being described below is related to a certain class of
{\it interactive information systems}. Namely, let us call interactive
information system {\it computer--graphic\/} if the information stream
from the computer is mounted as a stream of geometric graphical data on
a screen of the monitor; an interactive computer--graphic information system
is called {\it psychoinformation\/} one if the information from observer
to computer transmits unconsciousnessly. Below we shall consider only such
systems.

We shall define conceptes of the interpretational figure and its symbolic
drawings, which as it seems play a key role in the description of the
computer--geometric representation of mathematical data in such interactive
information systems.

Mathematical data in interactive information system exist in the form of an
interrelation of {\it an interior geometric image\/} ({\it figure\/}) in the
subjective space of observer and {\it an exterior computer--graphic
representation}. The exterior computer--graphic representation includes
{\it the visible elements\/} (drawings of figure) as well as of {\it the invisible
ones\/} (f.e. analytic expressions and algorythms of the constructing of such
drawings). Process of the corresponding of a geometrical image (figure) in the
interior space of observer to a computer--graphic representation (visible and
invisible elements) will be called {\it translation}. For example, a circle as
a figure is a result of the translation of its drawing on a screen of the
videocomputer (the visible object), constructed by an analytic formula (the
invisible object) accordingly to the fixed algorythm (also the invisible object).
It should be mentioned that the visible object may be nonidentical to the
figure, f.e. if a 3--dimensional body is defined by an axonometry, in three
projections, cross-sections or cuts, or in the window technique, which allows
to scale up a concrete detail of a drawing, etc; in this case partial visible
elements may be regarded as {\it modules}, which translation is realized
separately. We shall call the translation by {\it interpretation\/} if the
translation of partial modules is realized depending on the result of the
translation of preceeding ones and by {\it compilation\/} otherwise.
An example of the interpretation may be produced by the drawing of a fractal
which structure is defined by an observer on each step of the scaling up in
the window technique; the translation of visible elements in an intentional
anomalous virtual reality (see below) is also an interpretation.

\definition{Definition 1} A figure, which is obtained as a result of the
interpretation, will be called {\it interpretational figure\/}.
\enddefinition

It should be mentioned that an interpretational figure may have no any
habitual formal definition; namely, only if the process of interpretation has
an equivalent compilation process then the definition of figure is reduced to
sum of definitions of its drawings; nevertheless, in interactive information
systems not each interpretation process has an equivalent compilation one.
It means that an interpretational figure has no any finite or constructively
determined set of basic properties, from which other properties are derived in
a formally logical way.

Note also that the drawing of interpretational figure may be characterized only
as "visual perception technology" of figure but not as an "image", such drawings
will be called {\it symbolic\/}.

The computer--geometric description of mathematical data in interactive
information systems is deeply related to the concept of anomalous virtual
reality.

It should be mentioned that there exist not less than two approaches to
foundations of geometry: in the first one the basic geometric objects are
figures defined by their drawings, geometry describes realtions between them, in
the second one the basic geometric concept is a space (a medium, a field),
geometry describes various properties of a space and its states, which are
called the drawings of figures.

For the purposes of the describing of geometry of interactive information
systems it is convenient to follow the second approach; the role of the medium
is played by an anomalous virtual reality, the drawings of figures are its
certain states.

\subhead 1.2. Anomalous virtual realities \endsubhead

\definition{Definition 2}

A. {\it Anomalous virtual reality\/} ({\it AVR\/}) {\it in a narrow sense\/}
is a certain system of rules of non--standard descriptive geometry adopted to
a realization on videocomputer (or multisensor system of "virtual reality").
{\it Anomalous virtual reality in a wide sense\/} contains also an image in
the cyberspace made accordingly to such system of rules.
We shall use this term in a narrow sense below.

B. {\it Naturalization\/} is the corresponding of an anomalous virtual reality
to an abstract geometry or a physical model. We shall say that the anomalous
virtual reality {\it naturalizes\/} the model and such model {\it
transcendizes\/} the naturalizing anomalous virtual reality. {\it
Visualization in a narrow sense\/} is the corresponding of certain images or
visual dynamics in the anomalous virtual reality to objects of the abstract
geometry or processes in the physical model. {\it Visualisation in a wide
sense\/} also includes the preceeding naturalization.

C. An anomalous virtual reality, whose images depends on an observer, is
called {\it intentional anomalous virtual reality\/} ({\it IAVR\/}).
Generalized perspective laws in IAVR contain the equations of dynamics of
observed images besides standard (geometric) perspactive laws. A process of
observation in IAVR contains a physical process of observation and a virtual
process of intention, which directs an evolution of images accordingly to
dynamical laws of perspective.
\enddefinition

In the intentional anomalous virtual reality objects of observation present
themselves being connected with observer, who acting on them in some way,
determines, fixes their observed states, so an object is thought as a
potentiality of a state from the defined spectrum, but its realization depends
also on observer.

The symbolic drawings of interpretational figures are presented by states of a
certain intentional anomalous virtual reality.

\subhead 1.3. Colors in anomalous virtual realities \endsubhead

It should be mentioned that the deep difference of descriptive geometry of
computer--graphic information systems from the classical one is the presense
of colors as important bearers of visual information. The reduction to shape
graphics, which is adopted in standard descriptive geometry, is very
inconvenient, since the use of colors is very familiar in the scientific
visualization. The approach to the computer--graphic interactive information
systems based on the concept of anomalous virtual reality allows to consider
an investigation of structure of a color space as a rather pithy problem of
descriptive geometry, because such space may be much larger than the usual one
and its structure may be rather complicated. Also it should be mentioned that
the using of other color spaces allows to transmit diverse information in
different forms, so an investigation of the information transmission via
anomalous virtual realities, which character deeply depends on a structure of
color space, become also an important mathematical problem.

\definition{Definition 3} A set of continuously distributed visual
characteristics of image in an anomalous virtual reality is called {\it
anomalous color space\/}. Elements of an anomalous color space, which have
non--color nature, are called {\it overcolors\/}, and quantities, which
transcendize them in the abstract model, are called {\it "latent lights"}.
{\it Color--perspective system\/} is a fixed set of generalized perspective
laws in fixed anomalous color space.
\enddefinition

\subhead 1.4. Mobilevision \endsubhead

{\it Mobilevision\/}  may be defined as a certain anomalous virtual reality,
which naturalizes the so--called {\it quantum projective field theory} [1].
However, here we prefer to explicate such definition in more technical terms.

\definition{Definition 4} {\it Mobilevision\/} is an artificial
computer--graphic interactive psychoinformation system with a projective
invariant feedback determined by eye motions of observer.
\enddefinition

Let's discuss this definition.
First, mobilevision is an artificial interactive information system (this point
corresponds to term "virtual" in the first form of the definition). Principles
of its construction are self--consistent and do not copy automatically any
natural laws just like {\it principles of airplane's construction differs from
ones of bird's physiology\/} (this point corresponds to the term "anomalous").
So mobilevision tries, first of all, to be a useful informatic construction but not
a model of any (may be rather important) natural phenomena.
Second, mobilevision is a computer--graphic information system, so an
information stream from a computer to a human is mounted in a form of images on
the screen; also it is a dynamical interactive system, i.e. the computer
changes geometric data on the screen by a certain algorythm, and such changes
depend on a behaviour of observer. Third, mobilevision is a psychoinformation
interactive system, i.e. characteristics of human behaviour, which are
available to the computer, have a subconscious character.
Fourth,
mobilevision is a very special psychoinformation system, a core of the
subconscious information stream from a human to a computer is geometric,
namely, consists of geometric data on eye motions. Such data may be reduced to
the coordinates of a sight point on the screen and its velocity.
Fifth, because both information streams in the mobilevision interactive system are
essentially geometric, there is postulated a geometric correlation between them.
Such correlation is encapsulated in dynamical laws of images realized by a
certain algorythm. These law should be projectively invariant with respect to
simultaneous projective transformations of image and sight geometric data.
However, a self--evident claim of projective invariance does not specify the
dynamical laws completely. Another invariance of dynamical laws is related to
symmetries of a color space. At the first approximation one has (due to
Maxwell, Helmholtz and Young) a $\SUt$ color symmetry (see par.2.1.), which is
really, however, broken. Nevertheless, an approximate $\SUt$--symmetry is a
rather natural mathematical startpoint. Thus, one claims the dynamical laws of
mobilevision to be $\SUt$--invariant.

The described suppositions are sufficient for a mathematization of
mobilevision, i.e. for a derivation of the using of all necessary mathematical
requisites from the first principles of mobilevision.

First, let's represent all geometric continuously distributed data of image by
certain quantities $f_i(x,y)$, where $(x,y)$ are coordinates on the screen.
It is convenient to use their chiral factorisation $f_i(x,y)=\sum_{j,k}
a_{ijk}\phi_j(z)\phi_k(\bar z)$, where $\phi_j(z)$ are holomorphic functions of
a complex variable $z$. The projective group $\PSL(2,\Bbb C)$ (or, at least,
its Lie algebra $\sLtwo$) acts on the quantities $\phi_j(z)$ ("fields") as on
holomorphic $\lambda$--differentials. The color group $\SUt$ also acts on
them globally, i.e. transforms them by a rule independent on a point. Actions
of $\sLtwo$ and $\SUt$ commute.

Second, let $u$ be a complex coordinate of a sight point, $\dot u$ be its
velocity. It is rather natural to suppose that the dynamical laws are
differential and that they express the first time--derivatives of "fields" as
linear operators of "fields" themselves with coefficients depending on $u$ and
$\dot u$. The general form of such laws was written in [2]. The differential
equations were interpreted as quantum--field analogs of the Euler formulas.
It should be marked that a quantum--field meaning was given to these formulas
by their interpretation and was not derived from general invariance principles.
However, such interpretation is a useful source to pick out the most important
cases of the dynamical laws. However, one may avoid it and to have deal with
operators in dynamical laws in purely mathematical fashion as with the vertex
operator fields for the Lie algebra $\sLtwo$. Such vertex operator fields form
a certain algebraic structure (QPFT--operator algebra) described in details in
[3]. However, the dynamical differential equations possess also $\SUt$ color
symmetry, it manifests itself also as a symmetry of the related QPFT--operator
algebra. QPFT--operator algebras with additional $\SUt$--symmetries were
described in [3] under the title of projective $\SUt$--hypermultiplets.
The most natural class of projective $\SUt$--hypermultiplets (the canonical
projective $G$--hypermultiplets, $\SUt\subset G$) was considered.

However, solitary Euler formulas are not $\SUt$-invariant, so they should be
completed by any other formulas. The most natural way to complete classical
Euler formulas is to consider the Euler--Arnold equations. In our
"quantum--field" case it means to consider the operators of dynamical laws (of
the "quantum--field" Euler formulas) to be explicitely depending on a time,
and to postulate their evolution to be governed by the Euler--Arnold equations
[3]. The least have a hamiltonian form, and if a hamiltonian is
$\SUt$--invariant then the complete dynamical laws will be also
$\SUt$--invariant.

So the basic dynamical laws of mobilevision in a form of the "quantum--field"
Euler formulas coupled with the Euler--Arnold equations are derived from the
first principles. Note that "quantum--field" Euler formulas  are fixed uniquely
by the claim of projective invariance whereas the Euler--Arnold formulas may be
replaced by any other ones, which will also provide the dynamical laws by
$\SUt$--invariance. Nevertheless, the Euler--Arnold formulas are, indeed,
the most natural "anzatz".

Mark that mobilevision may be consider as a certain artificial form of
{\it interactive visions}, which exploration is of a strong perpetual interest.
F.e. it is rather intriguing to view this general topic in a context of multi--user
effects in interpretational geometries.

Let's discuss mobilevision dynamics once more (cf.[2,4]).

First, note that the eye motions are not homogeneous. One may extract three
different parts from them, namely, slow movements, saccads and tremor. The
least may be naturally stochastized, i.e. be simulated by a certain stochastic
process. It is resulted in an additional stochastic term in the Euler formulas.
However, one may consider Euler formulas with an additional term from the
beginning. In this case the dynamical laws are described by a stochastic linear
differential equation of the form $\dot\Phi=A(u,\dot u)\Phi\,dt+
B(u,\dot u)\Phi\, d\omega$, where operator fields $A$ and $B$ are independent
(certainly, such equations are coupled with the deterministic Euler--Arnold
equations on $A$ to provide $\SUt$--invariance). To maintain the
$\SUt$--invariance one should claim $B$ to be $\SUt$--invariant. Therefore,
the most natural anzatz is to relate $B$ to a $\SUt$--invariant spin--1
vertex operator field (current) in the projective $\SUt$--hypermultiplet.
The resulted stochastic equations are formally a certain "quantum--field"
analog of a form of Belavkin equations but without Belavkin counterterm, which
provides exceptional nondemolition properties for solutions of Belavkin
equations. Because this is the useful effect, we may include a "quantum--field"
analog of Belavkin counterterm (determined by a spin--2 $\SUt$--invariant
vertex operator field) in our equations by hands.

\head 2. Octonionic binocular mobilevision
\endhead

\subhead 2.1. Quaternionic description of ordinary color space \endsubhead

It should be mentioned that ordinary color space may be described by use of
imaginary quaternions in the following way: let us consider an arbitrary
imaginary complex quaternion $q=ri+bj+gk$, $i, j, k$ are imaginary roots and
$r, g, b$ are complex numbers.
One may correspond to such quaternion an element of the color space, which in
RGB--coordinates has components $R=|r|^2$, $G=|g|^2$, $B=|b|^2$.
The lightening $L$ has the quadratic form in the quaternionic space, namely,
$L=\frac12(|r|^2+|b|^2+|g|^2)$.
The group $\SUt$ is a group of its invariance.

One may also consider any other coordinate systems XYZ in the color space
such that RGB are linear combinations of XYZ. Then, the quaternion has the
form $q=xi+yj+zk$ and the lightening $L$ is proportional to $|x|^2+|y|^2+|z|^2$
(cf.[5]).

\subhead 2.2. Octonionic color space and binocular mobilevision \endsubhead

Let us construct an octonionic color space to describe the binocular
mobilevision.
This space will be a semi--direct product of a canonical projective
$\G_2$--hypermultiplet on the trivial one, which is a direct sum of seven
copies of the suitable Verma module over $\slt$.
The group $\G_2$ acts in this seven dimensional space as it acts on imaginary
octonions [6].
There is uniquely defined up to a multiple and modulo the Poissonic center an
$\SUt$--invariant quadratic element in $S^2(\frak g_2)$.
So we can construct the Euler--Arnold equations in the canonical projective
$\G_2$--hypermultiplet.
To receive the binocular version of the affine Euler formulas one should use
the decomposition of $S^2(\frak g_2)$ on the $\SUt$--chiral components (left
and right); the angular fields from the chiral components will depend on
chiral parameters $u_l$, $\dot u_l$ and $u_r$, $\dot u_r$,
attributed to the left and right eyes, respectively.
Six copies of Verma modules over $\slt$, mentioned above, form a pair of
projective $\SUt$--hypermultiplets, which correspond to ordinary colors for
left and right eyes; one copy form also a projective $\SUt$--hypermultiplet,
its overcolor will be called {\it a strange overcolor}.
So the constructed seven dimensional octonionic color space includes a pair
of ordinary three dimensional color spaces (for left and right eyes,
respectively) and one strange overcolor.

Binocular mobilevision may be realised on any PC by use of two special
components: (1) stereo glasses (f.e. 3Dmax of Kasan Electronics Co., Ltd.),
and (2) any computer system of the real--time biomedical data acquisition.
One may use dynamical perspective laws different from described above. Really,
it is very convenient to use methods of dynamical interactive screening
(analogous to noninteractive dynamical screening of 3Dmax).

\Refs
\roster
\item"[1]" Juriev, D.V., Octonions and binocular mobilevision. E--print
(LANL Electronic Archive on Theor. High Energy Phys.): {\it hep-th/9401047\/}
(1994).
\item"[2]" Juriev, D.V., {\it Theor. Math. Phys.\/} 92(1) (1992) 814-816.
\item"[3]" Juriev, D.V., {\it Theor. Math. Phys.\/} 98(2) (1994) 147-161.
\item"[4]" Juriev, D.V., {\it Theor. Math. Phys.\/} 106(2) (1996).
\item"[5]" Leonov, Yu.P., In {\it Color problem in psychology}. Moscow, Nauka,
1993, p.54-67; {\it J. Psychol. (Russian)\/} 16(2) (1995) 133-141.
\item"[6]" Freudenthal H., {\it Geom. Dedicata\/} 19 (1985) 7-63.
\endroster
\endRefs
\enddocument